\documentstyle[12pt,epsf]{article}
\def\lsim{\mathrel{\lower2.5pt\vbox{\lineskip=0pt\baselineskip=0pt
           \hbox{$<$}\hbox{$\sim$}}}}
\def\gsim{\mathrel{\lower2.5pt\vbox{\lineskip=0pt\baselineskip=0pt
           \hbox{$>$}\hbox{$\sim$}}}}
\begin{document}
\setlength{\baselineskip}{8mm}
\begin{titlepage}
\begin{flushright}
\begin{tabular}{c c}
& {\normalsize  hep-ph/9604362} \\
& {\normalsize  DPNU-96-22} \\
& {\normalsize April 1996}
\end{tabular}
\end{flushright}
\vspace{5mm}
\begin{center}
{\large \bf Large Majorana Mass  \\ 
from the Anomalous $U(1)$ Gauge Symmetry} \\
\vspace{15mm} 
Naoyuki Haba\footnote{E-mail:\ haba@eken.phys.nagoya-u.ac.jp} \\ 

{\it 
Department of Physics, Nagoya University \\
           Nagoya, JAPAN 464-01 \\
}
\end{center}

\vspace{10mm}

%
\begin{abstract}
We show a new and simple model which induces large 
Majorana masses of right-handed neutrinos. 
It is based on $U(1)_X$ anomalous gauge symmetry, which 
is cancelled by the Green-Schwarz mechanism. 
These Majorana masses can solve the solar 
neutrino problem by the vacuum oscillation mechanism. 
The superpotential of this model is scale 
invariant. 
The field contents are simple extensions of the 
MSSM, which have four 
standard gauge singlet fields, 
two extra vector-like fields, and right-handed neutrinos 
in addition to the MSSM. 
\end{abstract}
\end{titlepage}
%
%
\section{Introduction}
The SUSY theory which guarantees the smallness 
of scalar masses is the most 
attractive candidate beyond the standard model.
It is likely that the SUSY theory is an effective theory 
of string. 
In this paper, we investigate the possibility 
that such a theory can accommodate the solar 
neutrino problem. 
The solar neutrino problem suggests the existence 
of Majorana neutrinos with masses of 
$O(10^{10} \sim 10^{12})$ GeV for 
the Mikheyev-Smirnov-Wolfenstein (MSW) mechanism, 
or $O(10^{12} \sim 10^{14})$ GeV for the vacuum oscillation 
(VO) mechanism\cite{see-saw1}\cite{see-saw2}\cite{see-saw3}.
Then how can we understand these mass scales 
within the framework of the 
string inspired SUSY theory? 
We take here the string inspired model which has no scales 
except for the string scale and 
soft SUSY breaking terms. 
In this framework, 
it is difficult to build the model 
which can derive large Majorana masses. 
Although some models have succeeded to produce such an 
intermediate scale 
by using non-renormalizable interactions\cite{Haba}, 
we show here another simple model in which the superpotential 
have no dimensionful parameters. 
The superpotential is scale invariant, that is, renormalizable and 
composed of only cubic interactions. 
We know that the cubic coupling of the superpotential 
is calculable in some string theory. 
So we can search for a string compactification 
that might lead to this type of models. 
In the model proposed here, 
the anomalous $U(1)_X$ gauge symmetry is crucial in 
generating suitable Majorana masses. 
The $U(1)_X$ gauge anomaly is cancelled by 
the Green-Schwarz mechanism\cite{GS}. 
This model can naturally solve the solar neutrino problem 
by the VO mechanism. 
\par
We introduce four singlet fields, 
two vector-like extra generations, and right-handed neutrinos 
in addition to usual spectrum of the MSSM. 
These additional fields play following roles: 
\begin{enumerate}
\item Four singlet fields are introduced to make the 
superpotential scale invariant and generate large Majorana masses. 
By the suitable $U(1)_X$ charge assignment, 
we will show that 
only one of them acquire desired vacuum expectation 
value (VEV) and other three singlet fields get 
weak (SUSY breaking) scale VEVs. 
The former generates the Majorana masses and 
the latter give the so-called $\mu$ term and 
masses of extra generations. 
\item Two vector-like extra generations are 
needed to satisfy the condition of the Green-Schwarz mechanism. 
Unless there are fields 
which have color and/or EM charges, 
we have no solution for the Green-Schwarz mechanism. 
The existence of extra vector-like generations is the 
general feature in string theories, 
where the anomaly cancellation of 
$SU(3)_c \times SU(2)_L \times U(1)_{Y}$ 
is automatic. 
\item Right-handed neutrinos are added to solve the solar neutrino 
problem by the see-saw mechanism. 
\end{enumerate}
\par
Section 2 is devoted to the explanation of the 
anomalous $U(1)_X$ gauge symmetry. 
In section 3, we study the new model. 
Section 4 gives summary and discussion. 
%
%
\section{Anomalous $U(1)_X$ Symmetry}
If there is the anomalous $U(1)_X$ symmetry, 
a Fayet-Iliopoulos 
term is induced from the string 
loop effects\cite{FI}. 
In this case, $D$-term becomes 
\begin{equation}
\label{FIterm}
D = {g_S^2 \: M_S^2 \over 192 \: \pi^2}\; {\rm Tr} \: q \: + 
      \: \sum_{i} \: \varepsilon_i^{-1}\: q_i \: |S_i|^2 \:,
\end{equation}
where $g_S$ and $q_i$ are the string 
coupling, and the charge of the 
anomalous $U(1)_X$ gauge symmetry, 
respectively. 
$S_i$ is the scalar component of 
the $i$th superfield. 
$M_S$ is the string scale\cite{MS} which is given by 
\begin{equation}
\label{MS}
M_S \simeq 5.27 \times g_S \times 10^{17} \: {\rm GeV} .
\end{equation}
$\varepsilon_i$ is the small parameter which can be derived from 
moduli parameters as shown in the next section. 
\par
The scalar potential with soft breaking 
mass parameters becomes 
\begin{equation}
\label{Vscalar}
V = {1 \over 2}\: D^2 + \sum_{i} m_{i}^2 \: |S_i|^2 .
\end{equation}
Here we neglect $F$ terms and other 
soft breaking terms. 
We will see that it is justified 
for the model shown in the next 
section. 
The stationary condition for $S_i$ is 
\begin{equation}
\label{station}
{\partial V \over \partial S_i} = \: D \: q_i S_i^* + 
                                    m_i^2 S_i^* = 0 .
\end{equation}
If $S_i^*$ gets the non-zero VEV, 
we obtain the relation 
\begin{equation}
\label{D}
D = - {m_i^2 \over q_i} .
\end{equation}
Note that this requires 
\begin{equation}
\label{softm}
m_i^2 \propto q_i . 
\end{equation}
It is clear that Eq.(\ref{softm}) cannot 
be satisfied for an arbitrary $U(1)_X$ 
charge. 
Then, we can conclude that the only $i$th field 
denoted by $\phi$ has the large VEV. 
The value of VEV becomes 
\begin{equation}
\label{VEV}
\langle \phi \rangle 
= \sqrt{\left( -{g_S^2 \over 192 \pi^2} 
    \; M_S^2 \; {\rm Tr}\; q  
 - {m_i^2 \over q_{\phi}} \right){\varepsilon_{\phi}\over q_{\phi}}} 
\simeq \sqrt{-{g_S^2 \over 192 \pi^2} 
             {{\rm Tr}\; q \over q_{\phi}}\varepsilon_{\phi}}\; M_S 
\end{equation}
from Eq.(\ref{D}). 
In the next section, we will show 
that this VEV yields 
Majorana masses of right-handed neutrinos. 
%
%
\section{A Simple Model}
Now we consider a simple model which 
has large Majorana masses of right-handed 
neutrinos. 
This model is renormalizable in contrast to 
the models proposed in Ref.\cite{hiera}. 
The superpotential $W$ has only cubic interactions. 
It assures us of using Eq.(\ref{Vscalar}) for the 
scalar potential because we assume that squarks and sleptons donot 
obtain VEVs. 
\par
$W$ is written as 
\begin{eqnarray}
\label{W}
 W &=&   y_{ia}^e L_i H_1 E^c_a + 
         y_{ij}^n L_i H_2 N^c_j + 
         y_{ij}^d Q_i H_1 D^c_j + 
         y_{ia}^u Q_i H_2 U^c_a   \nonumber  \\
   & & + f_{ij}^{(1)} \phi N^c_i N^c_j + 
         f^{(2)} X H_1 H_2 + 
         f_a^{(3)} Y \overline{U^c} U^c_a + 
         f_a^{(4)} Z \overline{E^c} E^c_a  \; , 
\end{eqnarray}
where $Q_i, L_i, D_i, N_i$, and $H_{1,2}$ are 
quark doublets, lepton doublets, right-handed 
down-sector quarks, right-handed neutral leptons, 
and Higgs doublets, respectively. 
$X, Y, Z$, and $\phi$ are standard gauge singlet fields. 
The indices $i, j$, and $a$ denote 
numbers of generations, which satisfy 
$i, j = 1 \sim 3$ and $a = 1 \sim 4$. 
There exist the fourth generation and mirror fields 
denoted by 
$\overline{U^c}$ and $\overline{E^c}$ for 
right-handed up-sector quarks $U^c$ and 
right-handed charged leptons $E^c$, respectively. 
By introducing these extra vector-like fields, 
we can get useful models which 
have rich $CP$ structures\cite{CP}. 
As we said before, we assume 
that only fields 
$X, Y, Z, H_{1,2}$, and $\phi$ 
get VEVs in order not to break 
color and EM charge, and in order to 
keep SUSY unbroken down to the weak scale. 
Then the $f^{(2)}$ coupling only contribute to the 
stationary condition in Eq.(\ref{station}) for the 
$F$ term. 
This $F$ term effect can not let the 
$j$th field $(i \neq j)$ satisfy 
Eq.(\ref{softm}) without unnatural fine-tuning. 
So we can say that if the field has a distinct 
$U(1)_X$ charge, only $\langle \phi \rangle$ is large and 
$\langle X \rangle$, 
$\langle Y \rangle$, $\langle Z \rangle$, and 
$\langle H_{1,2} \rangle$ are of 
order weak (SUSY breaking) scale. 
The suitable $\mu$ term is effectively 
derived from $f^{(2)} \langle X \rangle$\cite{NMSSM}. 
\par
The mixed anomaly coefficients for 
$U(1)_X$ and $SU(3)_c \times SU(2)_L \times U(1)_Y$ are given by  
\begin{eqnarray}
\label{C1}
C_1 &=& {1 \over 6} ( 3 [q_{H_1} + q_{H_2}] + 
        3 [q_{Q} + 8 q_{U^c} + 2 q_{D^c} + 
        3 q_{L} + 6 q_{E^c}]               \nonumber \\
    & & + 8 q_{U^c} + 8 q_{\overline{U^c}} + 
        6 q_{E^c} + 6 q_{\overline{E^c}} ), \\
\label{C2}
C_2 &=& {1 \over 2} \left( q_{H_1} + q_{H_2} + 
        3 [ 3 q_{Q} + q_{L}] \right), \\ 
\label{C3}
C_3 &=& {1 \over 2} \left( 
        3 [2 q_{Q} + q_{U^c} + q_{D^c}] + 
        q_{U^c} + q_{\overline{U^c}} \right), \\ 
\label{Cg}
C_{grav} &=& {1 \over 24} {\rm Tr} \; q .
\end{eqnarray}
Here, $C_1, C_2$, and $C_3$ are the 
coefficients of the mixed 
$U(1)_X \times U(1)_Y^2$, 
$U(1)_X \times SU(2)_L^2$, and 
$U(1)_X \times SU(3)_c^2$ 
anomalies, respectively. 
$C_{grav}$ is the gravitational 
anomaly mixed with $U(1)_X$. 
The Green-Schwarz mechanism for 
anomaly cancellation requires 
\begin{equation}
\label{GSmecha}
{C_1 \over k_1} = {C_2 \over k_2} = 
{C_3 \over k_3} = {C_{grav} \over k_{grav}} .
\end{equation}
Where, $k_i$ denotes the Kac-Moody level, 
which satisfies $k_1 = 5/3$, 
$k_2 = k_3 = k_{grav} = 1$. 
More general case is argued in Ref.\cite{Ibanez}. 
The $U(1)_X^2 \times U(1)_Y$ anomaly 
denoted by $C_{XXY}$ cannot 
be cancelled by the Green-Schwarz mechanism. 
Then the equation  
\begin{eqnarray}
\label{GXXY=0}
C_{XXY} &=& - q_{H_1}^2 + q_{H_2}^2 + 
          3 [q_{Q}^2 - 2 q_{U^c}^2 + q_{D^c}^2 - 
          q_{L}^2 + q_{E^c}^2 ]      \nonumber \\
        & &  - 2 q_{U^c}^2 + 
          2 q_{\overline{U^c}}^2 + q_{E^c}^2 - 
          q_{\overline{E^c}}^2  \nonumber \\
        &=& 0 
\end{eqnarray}
must be satisfied to the anomaly. 
As for $C_{XXX}$, there is a possibility 
that other unknown particles 
may contribute. 
So we donot consider it. 
\par
There are fourteen independent $U(1)_X$ charge parameters. 
However, we want to consider only cubic interactions for $W$ 
as Eq.(\ref{W}), these charge parameters are 
reduced to six independent parameters by 
using the eight relations as 
\begin{eqnarray}
\label{8relations}
 & & q_{\phi} + 2  q_{N^c} = 0,        \nonumber \\
 & & q_{L} + q_{H_1} + q_{E^c} = 0,   \nonumber \\
 & & q_{L} + q_{H_2} + q_{N^c} = 0,   \nonumber \\
 & & q_{Q} + q_{H_1} + q_{D^c} = 0,   \nonumber \\
 & & q_{Q} + q_{H_2} + q_{U^c} = 0,  \\ 
 & & q_{X} + q_{H_1} + q_{H_2} = 0,   \nonumber \\
 & & q_{Y} + q_{U^c} + q_{\overline{U^c}} = 0,  \nonumber \\
 & & q_{Z} + q_{E^c} + q_{\overline{E^c}} = 0.  \nonumber 
\end{eqnarray}
Now we take $q_{H_1}, q_{H_2}, q_{Q}, q_{\overline{U^c}}, 
q_{\overline{E^c}}$, and $q_{N^c}$ for six 
independent parameters. 
Imposing these constraints, 
three equations of the Green-Schwarz 
mechanism for the anomaly cancellation 
Eq.(\ref{GSmecha}) becomes 
\begin{eqnarray}
\label{GSmecha2}
 & & 4 q_{H_1} + q_{H_2} + 10 q_{Q} - 3 q_{N^c} 
                        - q_{\overline{U^c}} = 0,         \\
\label{GSmecha3}
 & & 12 q_{H_1} - 6 q_{H_2} + 30 q_{Q} - 15 q_{N^c} 
         - 3 q_{\overline{U^c}} - 6 q_{\overline{E^c}} = 0,  \\ 
\label{GSmecha4}
 & & 25 q_{H_1} + 35 q_{H_2} + 10 q_{Q} - 2 q_{N^c} 
                        - 10 q_{\overline{U^c}} = 0.  
\end{eqnarray}
The equation (\ref{GXXY=0}) becomes to be 
\begin{eqnarray}
\label{GXXY02}
 C_{XXY} &=& -q_{H_1}^2 + q_{H_2}^2 + 
       3 q_{Q}^2 - 8 (q_{Q} + q_{H_2})^2 + 3 (q_{Q} + q_{H_1})^2 
                                                      \nonumber \\
 & & - 3 (q_{N^c} + q_{H_2})^2 + 4 (-q_{H_1} + q_{H_2} + q_{N^c})^2 
     + 2 q_{\overline{U^c}}^2 - q_{\overline{E^c}}^2 \nonumber \\
 &=&  0. 
\end{eqnarray}
{}From Eqs.(\ref{VEV}) and (\ref{W}), $f^{(1)}_{ij}$ terms 
induce supersymmetric Majorana mass terms. 
The value of Majorana mass $M_{M}$ becomes 
\begin{equation}
\label{Majo}
M_{M} \simeq \sqrt{-{g_S^2 \over 8 \pi^2} 
             {C_{grav} \over q_{\phi}}\varepsilon_{\phi}} \; M_S 
          \simeq \sqrt{-{1 \over 50 \pi} 
             {C_{grav} \over q_{\phi}}\varepsilon_{\phi}} \; M_S 
          = \sqrt{{1 \over 1200 \pi} \alpha \varepsilon_{\phi}}\; M_S \;\;,
\end{equation}
where, 
\begin{equation}
\label{alpha}
\alpha \equiv -{24 \; C_{grav} \over q_{\phi}}. 
\end{equation}
We use $g_S^2/4 \pi \simeq g_{GUT}^2/4 \pi 
\simeq 1/25$ in Eq.(\ref{Majo}). 
In addition to the above four equations 
(\ref{GSmecha2})$\sim$(\ref{GXXY02}), 
we get one more equation from the positivity of 
$\alpha$, that is 
\begin{equation}
\label{alpha2}
13 q_{H_1} + 11 q_{H_2} + 2 q_{Q} + 2 q_{N^c}(1 + \alpha) 
           - 2 q_{\overline{U^c}} = 0, \;\;\;\;\; \alpha > 0. 
\end{equation}
\par
Now we get the system of six unknown parameters 
and five equations. 
We can easily solve these five homogeneous equations. 
The result is that all 
$U(1)_X$ charges can 
be parameterized by $\alpha$. 
The ratio of the charges of $H_{1,2}$ becomes 
\begin{equation}
\label{h1/h2}
{q_{H_2} \over q_{H_1}} = 
-{2375 \alpha^2 + 7900 \alpha + 5832 \over 
  2125 \alpha^2 + 7600 \alpha + 5832} \;\;,
\end{equation}
and other charges are obtained from 
Eqs.(\ref{8relations}) $\sim$ (\ref{GSmecha4}). 
In fact, there is another solution 
of Eq.(\ref{h1/h2}), which is $q_{H_2}/q_{H_1} = -1$. 
However, we cannot get a large Majorana 
mass because $q_{\phi} = 0$ in this case. 
\par
Now we are in a place to discuss about values of 
$\alpha$ and $\varepsilon_{\phi}$
\footnote{We thank Professor Y. Kawamura for the 
discussion about $\alpha$ and $\varepsilon_{\phi}$.}. 
As for $\alpha$, 
the massless condition induces $\alpha \geq O(10^{-2})$ 
in the $Z_N$ orbifold model\cite{KAWAMURA}. 
On the other hand, 
$\varepsilon_{\phi}$ might be written in terms of 
the moduli field $T$ and the modular weight $n$ of $\phi$ as 
\begin{equation}
\label{epsilon}
{\varepsilon_{\phi}} \simeq {1 \over (T + T^*)^n}.
\end{equation}
Here $T$ can be the order $O(10)$
\footnote{It is favorable for the string 
gauge unification\cite{unify1}\cite{unify2}.} 
and the modular weight $n$ can be positive
in the twisted sector of 
the orbifold model
\footnote{Since $\phi$ has the $U(1)_X$ charge, 
$n$ cannot be large integer in practice. 
However $n=1$ is really possible for 
the specific orbifold model as in Ref.\cite{unify2}.}. 
Then some orbifold models might really 
derive $\alpha \cdot \varepsilon_{\phi} \simeq 10^{-2}$. 
\par
In this case, 
the Majorana mass becomes 
$6.1 \times 10^{14}$ GeV from Eq.(\ref{Majo}). 
This is suitable for the VO mechanism solution 
of the solar neutrino problem. 
However it is impossible to obtain the MSW solution which demands 
$\alpha \cdot \varepsilon_{\phi} \simeq 10^{-6}$ 
in ordinary compactifications. 
%
%
\section{Summary and Discussion}
We have studied a new and simple 
model which can induce large Majorana masses of 
the right-handed neutrinos, 
which can solve the solar neutrino problem by 
the VO mechanism. 
There is no scale in the original superpotential. 
The mass scale of Majorana masses is generated 
dynamically. 
It is the result of 
anomalous $U(1)_X$ gauge symmetry, 
which are cancelled by the Green-Schwarz 
mechanism. 
By the assignment of $U(1)_X$ charges, 
we can obtain suitable Majorana masses for the 
solar neutrino problem. 
Do the situation change if 
we change the number of singlet fields and/or 
vector-like generations? 
In the case of decreasing the number of fields, 
we have no solutions of the Green-Schwarz 
mechanism. 
It is easily shown by constructing 
the model which contain only $D^c$ $(U^c, E^c, N^c, Q, L)$, 
$\overline{D^c}$ $(\overline{U^c}, \overline{E^c}, 
\overline{N^c}, \overline{Q}, \overline{L})$, and one singlet 
in addition to NMSSM with cubic terms. 
We have also no suitable solution 
in the next three cases; 
\begin{enumerate}
\item NMSSM $+ (D^c, \overline{D^c}) + 
(E^c, \overline{E^c})+$ two singlets + right-handed neutrinos. 
\item NMSSM $+ (D^c, \overline{D^c}) + 
(N^c, \overline{N^c})+$ two singlets + right-handed neutrinos.
\item NMSSM $+ (U^c, \overline{U^c}) + 
(N^c, \overline{N^c})+$ two singlets + right-handed neutrinos.
\end{enumerate}
As for the case of increasing the number of fields, 
the solution of the Green-Schwarz mechanism may 
become the simple integer charge of $U(1)_X$. 
It is worth searching whether 
this $U(1)_X$ charge assignment is derived 
from orbifold models or not. 

\vskip 1 cm
\noindent
{\bf Acknowledgements}\par
I would like to thank Professor Y. Kawamura for 
useful discussions and comments. 
I am also grateful to Professor A. I. Sanda and 
Professor T. Matsuoka for useful comments and 
careful reading of the manuscript.


\end{document}